\documentclass{article}

\usepackage[left=0.5in, right=0.5in, top=1in, bottom=1in]{geometry}
\usepackage[colorlinks]{hyperref}
\usepackage{amsmath}
\usepackage{amssymb}
\usepackage{newpxtext}
\usepackage{newpxmath}
\usepackage{orcidlink}
\usepackage[backend=biber, sorting=none]{biblatex}
\usepackage{graphicx}
\usepackage[font=small]{caption}

\definecolor{KUred}{RGB}{144, 26, 30}
\definecolor{randomblue}{HTML}{4b7aaf}

\hypersetup{citecolor=randomblue, linkcolor=KUred, urlcolor=black}

\DeclareCaptionLabelSeparator{colon}{\textcolor{KUred}{\textbf{{:\space}}}}
\DeclareCaptionLabelFormat{redbold}{\textcolor{KUred}{\textbf{#1 #2}}}
\begin{document}

\title{\vspace{-40pt}
A loser in both environments can survive by switching between them}

\author{
Riz Fernando Noronha \orcidlink{0009-0007-2923-3835}$^\dagger$, 
Kunihiko Kaneko \orcidlink{0000-0001-6400-8587},
}

\date{\textit{Niels Bohr Institute, University of Copenhagen, Copenhagen 2200, Denmark}\\ \vspace{5pt}
$^\dagger$ Corresponding author: \href{noronha@nbi.ku.dk}{noronha@nbi.ku.dk}}

\maketitle

\renewcommand{\abstractname}{\large Abstract\\}

\begin{abstract}
    \normalsize
    How can a species persist in an environment where it is always outcompeted? Using a minimal predator-prey model with environment-dependent parameters, we show that a predator driven to extinction in each of two static environments can survive indefinitely once the environment alternates between them fast enough. We derive the critical switching rate above which persistence occurs, and show that random (Poisson) switching needs to be faster than periodic switching in order to offset prolonged spells in the unfavorable environment. We then generalize the mechanism to any two-species system, and can predict persistence solely based on the sign of a single ``switching rescue function" assembled from the two boundary vector fields. This general result has broad reaching consequences: for instance, when applied to a standard model of viral dynamics, it predicts that two drugs which each clear a pathogen on their own can fail when alternated, giving a non-resistance-based explanation for the failure of drug-cycling strategies. Our results demonstrate that the tempo of environment change, as opposed to the environments themselves, can lead to species survival.
\end{abstract}

\vspace{40pt}

\begin{figure}[h]
    \centering
    \includegraphics[width=0.9\linewidth]{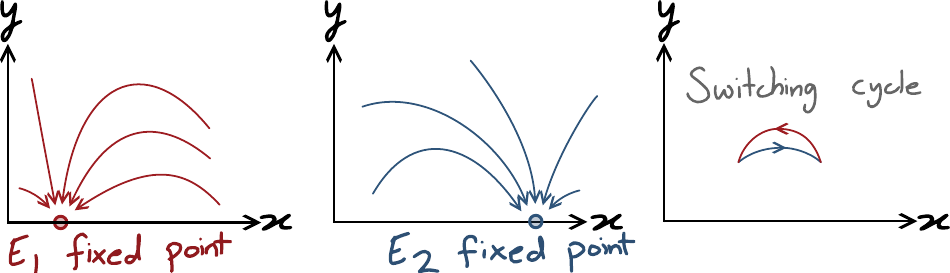}
    \caption{An illustration of the phase-space behavior. The two environments have different fixed points, both in which $y$ is eliminated, but the trajectories could be convex, leading to a transient where $y$ increases. Thus, switching quickly enough could lead to survival.}
    \label{fig:schematic}
\end{figure}

\twocolumn

\section{Introduction}

Ecosystems are comprised of a remarkable number of diverse species, coexisting and interacting with each other \cite{macarthur1984geographical, brown1981two}. While several arguments establish factors contributing to the coexistence, such as occupying different ecological niches \cite{adler2007niche}, or the role of spatial structure \cite{levin1974dispersion}, others focus on coexistence through environmental changes. Typically, such models consist of situations where different environmental conditions lead to different fit species, so that the environmental changes allow for coexistence \cite{hutchinson1961paradox, chesson1981environmental, chesson1994multispecies, chesson2000mechanisms}.

Here, we examine a different possibility: a species that is outcompeted in every static environment may potentially persist when the environment switches dynamically. In this way, we investigate the potential for coexistence inherent to a changing environment. Although the behavior has been documented in multiple recent models \cite{babajanyan2024microbial, hening2023random, wen2021alternating} and explored in a dynamical-systems framework for competitive systems \cite{Benaim2016}, a thorough investigation of the universal mechanism and the extent of its generality remains lacking.

The fundamental mechanism we propose is illustrated in \autoref{fig:schematic}. Suppose we have two species, $x$ and $y$, and two environments, $1$ and $2$. In both environments, the only fixed point is one where $y\!=\!0$, however they are distinct. If the route to the fixed point has $y$ increase transiently, switching between the two could keep it alive.

In this article, we explore the same behavior in a model which closely resembles standard predator-prey models used in population dynamics, and through dynamical systems theory, investigate the conditions under which such behavior emerges in a general two-species system.

\section{Model}

We create a simple two-species model as follows:

\begin{align}
    \frac{\mathrm{d}x}{\mathrm{d}t} &= r x\left(1-\frac{x}{K}\right) - x y  \label{eq:main-equation-dxdt} \\
    \frac{\mathrm{d}y}{\mathrm{d}t} &= x y - \gamma y \label{eq:main-equation-dydt}
\end{align}

The parameters $r$, $K$, and $\gamma$ represent the prey's growth rate, the prey's carrying capacity, and the predator's death rate respectively. These parameters are environment dependent.

In the case where $y\!=\!0$, the dynamics of $x$ correspond to a standard Verhulst or logistic equation \cite{verhulst1838notice}: it grows until it reaches a carrying capacity $K$, set by the environment. 

In the limit of $K\!\to\!\infty$, the model reduces to a Lotka-Volterra \cite{lotka1920analytical} system with biomass conservation, where all of the biomass consumed by the predator is used in its growth. The time is measured in units of the interaction term, and so the $xy$ term lacks a coefficient.

The model used resembles the Lotka-volterra with a time-varying carrying capacity of Swailem and Tauber \cite{swailem2023lotka}, however our parameters for the prey growth rate $r$ and predator death rate $\gamma$ are also allowed to vary under switching environments.

The standard parameters we use are $K_1\!=\!1, r_1\!=\!1, \gamma_1\!=\!1.1$ in environment 1, and $K_2\!=\!2, r_2\!=\!4, \gamma_2\!=\!2.1$ in environment 2. Environment 2 favours the prey more: it's growth rate and carrying capacity both increase, while the predator's death rate decreases.

\section{Results}

\subsection{Deterministic switching}

First, we study the switching between two environments at a fixed frequency. This corresponds to cases where the switching is periodic, such as entrainment to a circadian rhythm, seasonal cycles, and more. We characterize the switching by its rate $\alpha$: each environment is held for a time $1/\alpha$ before switching, so that one full period, returning to the same environment, takes $T=2/\alpha$.

Analytically, it is reasonable to consider two extremes: the slow switching case $\alpha\to0$, and the fast switching case $\alpha\to\infty$.

The slow switching case is identical to having a fixed environment. Here, we would like the system to be such that the predator $y$ dies out in both environments. To do so, we analyze the fixed points of the system. We observe that the predator is excluded if

\begin{equation}
    \gamma \geq K \label{eq:fixed-environment-death-inequality}
\end{equation}

For the fast-switching case, the dynamics are equivalent to a time-average of the two vector fields for each environment. We find the following condition under which fast-switching can rescue the system (see Materials and Methods for the derivation, as well the derivations for following results):

\begin{figure}[h!]
    \centering
    \includegraphics[width=\linewidth]{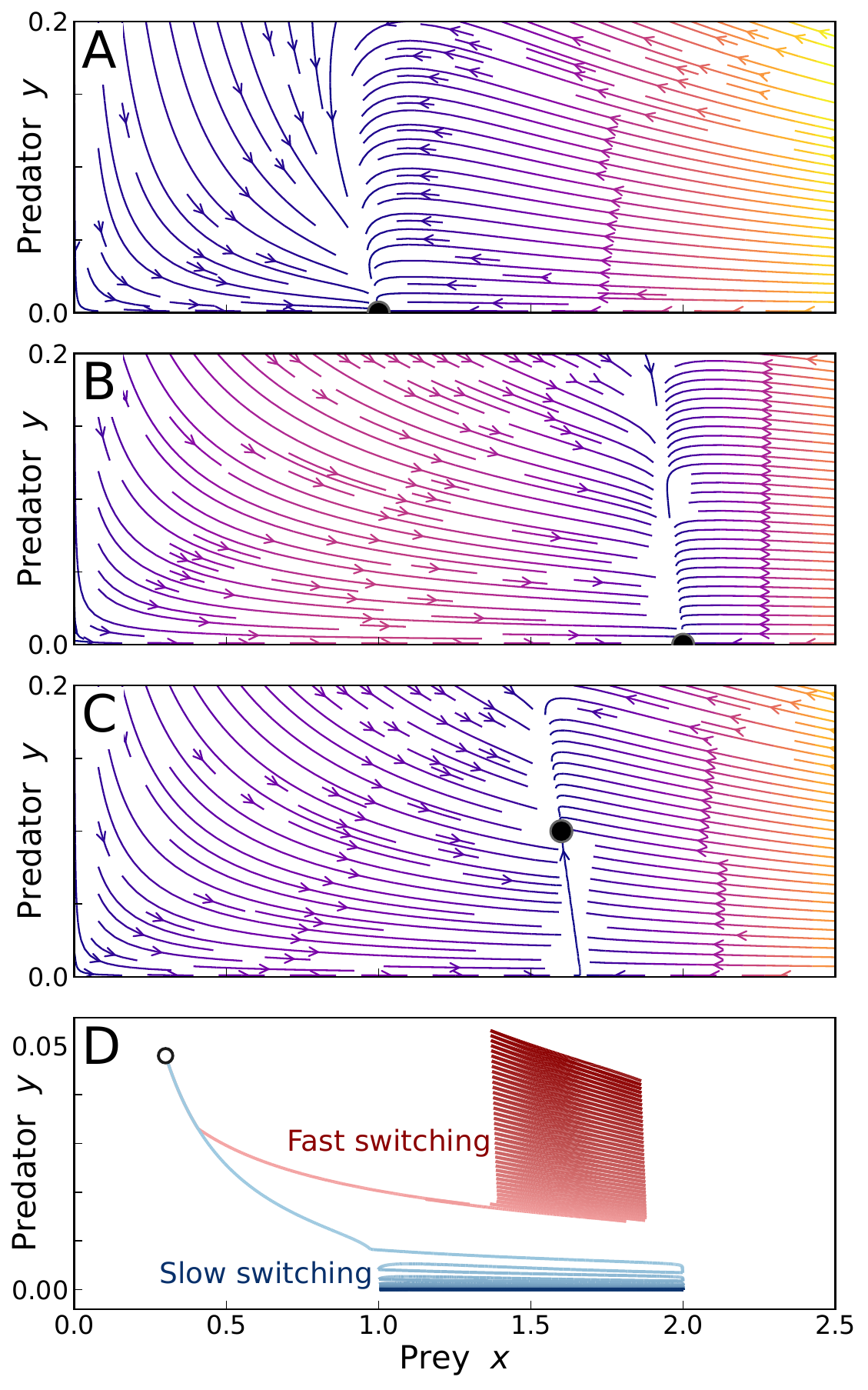}
    \caption{The dynamics of the model. A, B, C: Streamlines, where arrows indicate directions, and the color indicates the speeds. Black circles represent the fixed points. A: Environment 1, where $K_1\!=\!1, r_1\!=\!1, \gamma_1\!=\!1.1$; B: Environment 2, where $K_2\!=\!2, r_2\!=\!4, \gamma_2\!=\!2.1$; C: The averaged environment, equivalent to switching with $\alpha\!\to\!\infty$ between environments 1 and 2. The switching environment leads to a stable fixed point with $y^\star\!=\!0.1$, though $y^\star\!=\!0$ in the individual environments. D: trajectories of an initial condition ($0.3, 0.048$) under fast ($\alpha\!=\!2$) switching (red) which approaches a stable limit cycle, as well as slow ($\alpha\!=\!0.2$) switching (blue) which cycles at $y=0$.}
    \label{fig:vector-fields}
\end{figure}

\begin{align}
    (r_1 + r_2) - \left(\frac{r_1 \left( \gamma_1 + \gamma_2\right)}{2K_1} + \frac{r_2 \left( \gamma_1 + \gamma_2\right)}{2K_2}\right) > 0 \label{eq:switching-environment-persistence-inequality}
\end{align}

We can combine the two inequalities (Equations \ref{eq:fixed-environment-death-inequality} and \ref{eq:switching-environment-persistence-inequality}) to get the following sufficient and necessary conditions under which switching induces persistence:

\begin{equation} \label{eq:sufficient-necessary-condition1}
    \gamma_1 \geq K_1,\quad \gamma_2 \geq K_2
\end{equation}
\begin{equation} \label{eq:sufficient-necessary-condition2}
    \frac{\gamma_1 + \gamma_2}{2} < \frac{r_1 + r_2}{\frac{r_1}{K_1} + \frac{r_2}{K_2}}
\end{equation}

\autoref{eq:sufficient-necessary-condition1} implies that the system dies out in both fixed environments. \autoref{eq:sufficient-necessary-condition2} implies that the system can stay stable when switching infinitely fast. Thus, any system that satisfied these conditions will have the predator die out as $\alpha\to0$, yet persist above a critical switching rate $\alpha > \alpha_c^\text{det}$. The dynamics of a system that satisfies these conditions is presented in \autoref{fig:vector-fields}\footnote{$\alpha_c$ needs to increase extremely quickly as the boundary to \autoref{eq:sufficient-necessary-condition2} is approached, see Supporting Information Figure S2.}.

Floquet theory \cite{floquet1883equations, klausmeier2008floquet} allows us to study finite-time switching between the environments. We analyze whether a small perturbation of $y$ will grow or decline, governed by the Floquet exponent $\lambda$. The critical switching rate $\alpha_c^\text{det}$ can then be calculated as the solution to the transcendental equation

\begin{equation} \label{eq:Tc-transcendental-eqn}
    \frac{K_1\!+\!K_2\!-\!\gamma_1\!-\!\gamma_2}{2} = \frac{\alpha_c^\text{det}}{2}\left(\frac{K_1} {r_1}\!-\!\frac{K_2}{r_2}\right) \ln\!\left(\frac{x_a}{x_b}\right) ,
\end{equation}
where $x_a$ and $x_b$ are the endpoints of the periodic cycle near the $y=0$ boundary, and depend on $\alpha_c^\text{det}$, as well as $\mathcal{X}_1(t; x_\text{in})$ and $\mathcal{X}_2(t; x_\text{in})$, the flows in environments 1 and 2 from initial condition $x_\text{in}$:

\begin{equation} \label{eq:Tc-periodicity-conditions-xaxb}
    x_a=\mathcal{X}_1\!\left(\frac{1}{\alpha_c^\text{det}};x_b\right) , \quad
    x_b=\mathcal{X}_2\!\left(\frac{1}{\alpha_c^\text{det}};x_a\right)
\end{equation}

\begin{equation} \label{eq:logistic-flow-boundary-closed-form}
     \mathcal{X}_E(t; x_{\mathrm{in}}) = \frac{K_E x_{\mathrm{in}} e^{r_E t}}{K_E + x_{\mathrm{in}}(e^{r_E t} - 1)}
\end{equation}

In the more general case where the system spends a fraction $p$ of its time in environment $1$ and $(1\!-\!p)$ of its time in environment $2$, the condition \autoref{eq:sufficient-necessary-condition2} is similar, replacing the averages of $\gamma$, $r$ and $r/K$ with weighted averages (weighted by $p$ and $(1\!-\!p)$). More details, including the optimal $p$ to maximize $y$ or minimize $\alpha_c$, are presented in the Supporting Information (Sections 1, 2, Figure S1).

\subsection{Poisson switching}
 
\begin{figure}[h!]
    \centering
    \includegraphics[width=\linewidth]{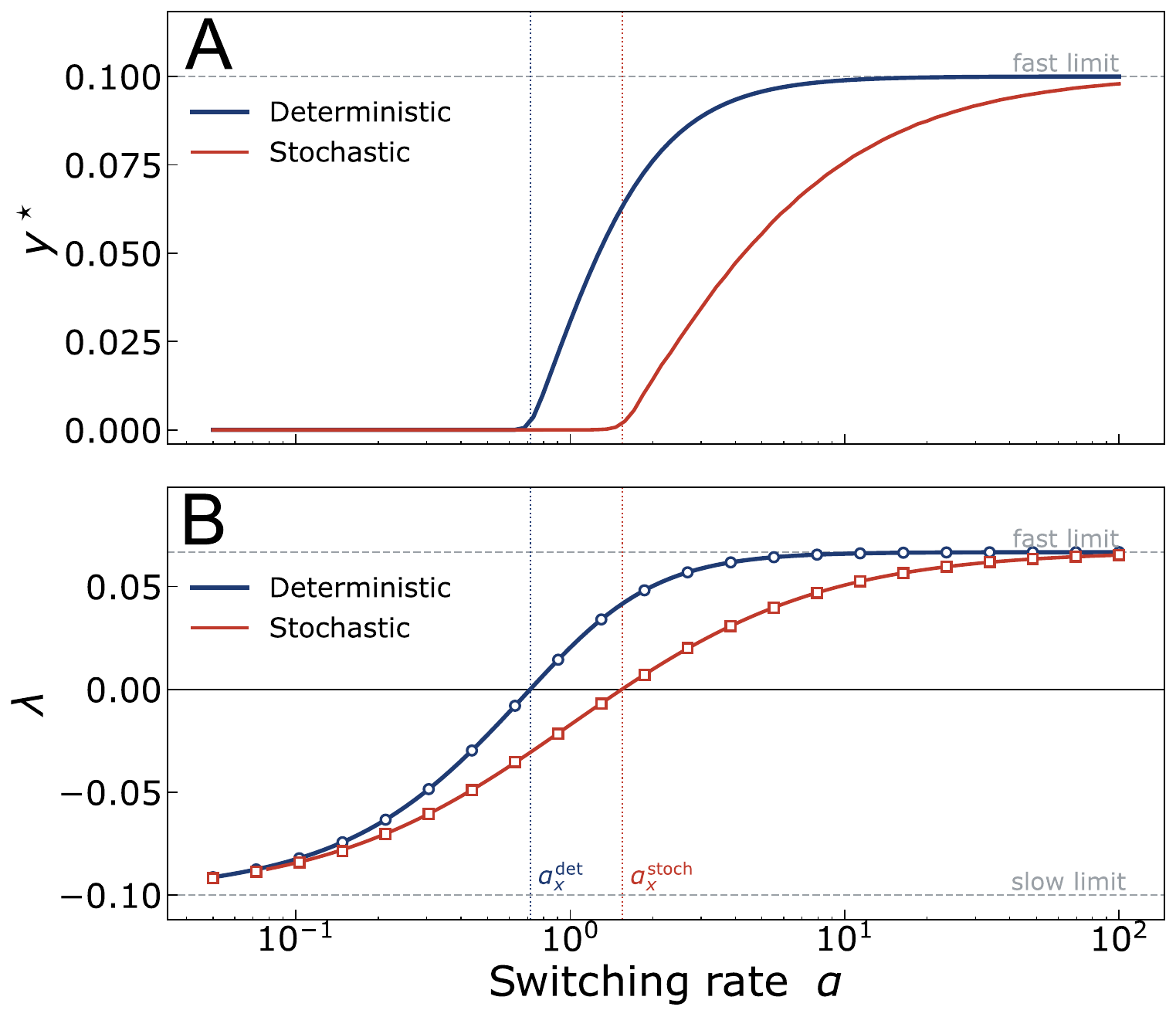}
    \caption{The mean predator population $y^\star$ (A) and the invasion exponent $\lambda$ (B) as a function of the switching rate $\alpha$. Stochastic switching needs to be faster, to compensate for the chance of getting `unlucky' and spending a long time in an environment. The curves represent Equations \ref{eq:Tc-transcendental-eqn} and \ref{eq:poisson-alphac-transcendental} (numerically solved for periodicity conditions on $x_a$ and $x_b$), which agree with the data from simulations (represented as markers).}
    \label{fig:ystar-lambda-vs-alpha}
\end{figure}
 
Here, we study the case where the switching is stochastic. This could represent environmental changes by irregular perturbations: for instance, rainfall. We consider the environmental change to be a Poisson process, where the environment switches with rate $\alpha$. On average each environment is occupied half the time, and so the time-averaged death rate is unchanged from the deterministic case.
 
As in the deterministic case, the predator persists above a critical switching rate $\alpha_c^\text{stoch}$\footnote{The `persistence' of the predator is only true on an infinite system. For finite populations, the predator could get unlucky and die even for $\alpha\!>\!\alpha_c$, as the only absorbing state is $y\!=\!0$.}, which is the solution to the transcendental equation
 
\begin{equation} \label{eq:poisson-alphac-transcendental}
    \frac{K_1\!+\!K_2\!-\!\gamma_1\!-\!\gamma_2}{2} = \frac{\alpha_c^\text{stoch}}{2}\left(\frac{K_1}{r_1}\!-\!\frac{K_2}{r_2}\right)\mathbb{E}\!\left[\ln\!\left(\frac{x_a}{x_b}\right)\right]
\end{equation}

This resembles \autoref{eq:Tc-transcendental-eqn}, with the deterministic log-ratio $\ln(x_a/x_b)$ replaced by its stationary average $\mathbb{E}\!\left[\ln\!\left(x_a/x_b\right)\right]$, the mean log-increment of the prey over a single environment-1 dwell, from $x_b$ at its start to $x_a$ at its end. Unlike the deterministic case, where $x_a$ and $x_b$ represent points and are algebraic, here $x_a$ and $x_b$ represent probability distributions of the turning points in the stationary state, $\mathbb{E}$ averages over both the exponential dwell time and the stationary distribution of $x_b$ at the start of the dwell.
 
Comparing the two thresholds, we find that stochastic switching must be faster, $\alpha_c^\text{stoch}\! >\! \alpha_c^\text{det}$ (\autoref{fig:ystar-lambda-vs-alpha}). The reason is that extra time in the favorable environment yields diminishing returns as the prey approaches its carrying capacity, whereas extra time in the damaging environment keeps costing the predator. A random sample could get unlucky, and be forced to undergo a long spell in the damaging environment. The predator must switch faster to compensate. For our standard parameters ($r_1\!=\!K_1\!=\!1$, $\gamma_1\!=\!1.1$, $r_2\!=\!4$, $K_2\!=\!2$, $\gamma_2\!=\!2.1$), we find $\alpha_c^\text{det}\!\approx\!0.716$ but $\alpha_c^\text{stoch}\!\approx\!1.55$, implying you need to switch twice as fast in the stochastic case.

\subsection{General framework}

We can consider the general case for a two-species system under a changing environment $i(t)\!\in\![1,2]$ to be

\begin{align}
    \frac{\mathrm{d}x}{\mathrm{d}t} &= F_{i}(x,y) \\
    \frac{\mathrm{d}y}{\mathrm{d}t} &= G_{i}(x,y)
\end{align}

In order for $F$ and $G$ to have species $y$ die out individually, we make the following assumptions:

\begin{enumerate}
    \item First, we assume that $y$ cannot recover from being $0$, for both environments i.e, $G_i(x,0) = 0$ for all $x \geq 0$. \label{general-assumption1}
    \item Next, we require that each environment has a fixed point at a given value $K_i$, i.e, $F_i(K_i, 0) = 0,\quad {F_i}^\prime(K_i, 0) < 0$. \label{general-assumption2}
    \item Finally, we assume there are no other stable manifolds in each environment, i.e, $F_i(x, y)=0$ and $G_i(x, y) = 0$ have no solutions besides $(K_i,0)$ in the first quadrant, and no limit cycles with $y\!>\!0$ exist. \label{general-assumption3}
\end{enumerate}

Without loss of generality, we label the environments based on the value of $x$ at their fixed points, i.e, $K_1 < K_2$. The conditions under which switching could cause persistence can then be represented by computing a `switching rescue function' $R(x)$, given by

\begin{equation}
    R(x) = \frac{\partial_y G_2(x,0)}{F_2(x,0)} - \frac{\partial_y G_1(x,0)}{F_1(x,0)} \label{eq:rescue-function}
\end{equation}

There exists a periodic switching cycle that can rescue species $y$ if and only if there exists an interval between $K_1$ and $K_2$ where $R(x)>0$. Conversely, if $R(x)\leq 0$ for all $x\in\left[K_1, K_2\right]$, there is no periodic switching that can lead to the persistence of $y$. 

\begin{figure}[h!]
    \centering
    \includegraphics[width=\linewidth]{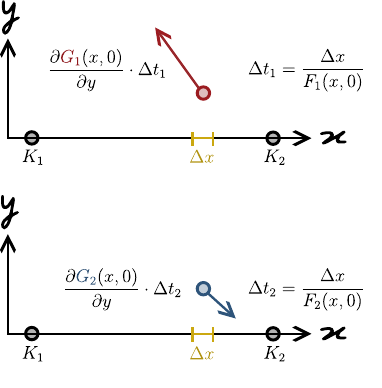}
    \caption{A schematic figure to explaining the condition under which switching induces persistence (\autoref{eq:rescue-function}). The `boost' in the predator population across a region $\Delta x$ is governed by the per-capita predator growth rate $\partial G_1/\partial y$, multiplied by the time spent in the region $\Delta t_1$. $R(x)$, defined as the sum across the two environments (environment $1$ on the top, environment $2$ below), tells us at what values of $x$ would a cycle transiently increase from the $y\!=\!0$ boundary, and thus whether the system can survive through switching.}
    \label{fig:Rfunction-expln}
\end{figure}

The rescue function $R(x)$ can be intuitively understood as follows (\autoref{fig:Rfunction-expln}): consider a small perturbation of $y$ at the boundary $y\!=\!0$. The per-capita growth rate, $\partial G(x,0)/\partial y$, depends whether it increases or decreases. However, the trajectory is changing $x$ due to the two environments having different fixed points, and thus the change in $y$ at a value of $x$ needs to be weighted by the time spent around that $x$ value, inversely proportional to $F(x, 0)$. Thus, \autoref{eq:rescue-function} is obtained from summing up these effects across both environments, and accounting for the change in signs due the $x$-trajectory being in opposing directions.

The general condition predicts the conditions where \textit{some} switching cycle can cause the system to survive, and thus while it cannot regenerate Equations \ref{eq:sufficient-necessary-condition1} and \ref{eq:sufficient-necessary-condition2}, it aligns with the conditions for survival if we vary the time spent in each condition. An illustration of the agreement between the region of predicted survival from $R(x)$ and the limit cycles for several models can be seen in the Supporting Information (Section 3, Figure S3).

Despite being extremely general, our condition is, in fact, conservative. The result on $R(x)$ holds even if there are other fixed points where $y\!>\!0$, as long as $y$ dies from a small perturbation. In other words, a system with a distant (unreachable) fixed point still has the same condition for switching to rescue $y$. Additionally, the equation only predicts the conditions where a small perturbation from $y\!\approx\!0$ can grow into a stable limit cycle: however switching-rescue could be more general by considering initial conditions (as there could exist a region of phase space where $G\!>\!0$).

Although the rescue function is derived for only two species and two environments, the latter is easily generalizable. If we have multiple environments, we simply need to calculate the pairwise $R(x)$ for each pair of environments. If even a single pair has a $R(x)\!>\!0$ for $x\!\in\![K_\text{min}, K_\text{max}]$, then some switching cycle between them can rescue $y$. If not, rescue is impossible.

\section{Discussion}

We identified the conditions under which a predator that's driven extinct in each of two static environments can persist once the environment switches between them. For periodic switching, persistence requires the switching rate to exceed a critical value $\alpha_c^\text{det}$, which can be derived implicitly. Stochastic switching can show the same behavior, but the irregularity of the dwell times forces the switching to be faster on average. Finally, we study a general case for any 2-variable system and find a simple criterion, the sign of a `switching rescue function' $R(x)$, which predicts if and when the phenomenon is possible.

Our mechanism is not the only one capable of creating this effect. For instance, an elegant paper by Lobry, Sciandra and Nival show that through using a separation of timescales and slow modes, a similar result can be obtained for competitive Lotka-Volterra interactions \cite{lobry1994effets}. However, the effect there is more intense, where $x$ goes extinct and the weaker species $y$ is the only one that persists, as opposed to a coexistence-based solution (expanded on by Benaim and Lobry \cite{Benaim2016}).

\subsection{Ecological implications}

The presented model is also abstract, not capturing several realistic details about population ecology. To the best of our knowledge, explicit experimental evidence of switching induced persistence remains unobserved. Nevertheless, our general analysis serves to present the conditions under which behavior emerges: conditions which we believe could occur in nature. 

There are two main ingredients required to observe switching induced persistence: the timescale of environmental switching should be faster than that of population dynamics \cite{li2016effects}, and the weaker species should obtain a transient population increase immediately after switching environments. Several macroscopic ecosystems have populations that take years to respond \cite{doney2013ecological}, yet seasons switch multiple times a year. A tradeoff mechanism between growth in a fixed environment and adaptation speed could give rise to a transient growth, but our assumption is much more general, and could arise by other means (such as interactions between the populations, as in the predator-prey model studied).

The same is true on the microscale. Recent models in microbial ecology highlight the need for changing environments \cite{abreu2020microbial}, showing how they could beat competitive exclusion \cite{bloxham2024biodiversity}. Although the doubling times of microbial species are extremely quick, the timescale at which species are excluded is often on the order of several days \cite{rodriguez2019rate, rodriguez2021rapid, nguyen2021environmental}. Thus, a daily switching of environments via sunlight, temperature, or circadian rhythms could be an unexplored factor for coexistence. Basan et al. demonstrate weaker strains of \textit{E. coli} are known to adapt faster, which could allow them to experience the transient growth boom \cite{basan2020universal}.

The current manuscript does not study spatial factors, which could be of interest regarding migration. Our general result also has the limitation of applying only to two-species systems. However, the concept could potentially be extended to $N$ species as well. If we consider the case where, in each environment, $N\!-\!1$ species survive, but $N$ can survive by switching, one could derive the invasion exponent in the dimension associated with the `losing' species, to see whether a perturbation collapses back to the $N\!-\!1$ dimensional manifold. There, however, could also be oscillatory solutions on the manifold, where our approach might not be easily applied.

\subsection{Applications to other fields}

The generality of our result has far-reaching implications. While we present it as a model for ecosystem dynamics, we believe that persistence through a fluctuating environment could be of relevance in several other areas in the biological sciences such as the dynamics of multi-host parasites, epidemiology, or oncology. As another example, performing the same analaysis on Nowak and May \cite{nowak2000virus} or Perelson's \cite{perelson2002modelling} model for viral dynamics reveals that drug cycling could potentially backfire (see Materials and Methods, \autoref{eq:nowak-may-perelson-conditions}): two different drugs which individually eliminate the pathogen could potentially fail when switching between them. While failure of drug cycling has been discussed in terms of evolving resistance \cite{levin2004cycling, bergstrom2004ecological}, our mechanism provides a different plausible explanation.

\subsection{Analogies}

In game theory, a related phenomenon is referred to as Parrondo's paradox \cite{harmer1999losing}, where an agent is able to win by switching between two losing games. The paradox has also been applied to several biological systems, mainly through bet-hedging-like mechanisms \cite{cheong2019paradoxical}. Parrondo's paradox has been applied, for example, to show how dormancy can be successful in predator prey-dynamics \cite{tan2020predator, wen2021alternating}.

While our phenomena itself has similarities to Parrondo's paradox \cite{harmer1999losing} (in which game B wins using the wealth distribution of game A, while in our case environment 1 can increase $y$ by using the initial condition of environment 2), there are a few differences. Firstly, our model is purely deterministic, while Parrondo's paradox is stochastic. Furthermore, in the canonical games of Parrondo's paradox, one wins on average through an ensemble distribution of wealths, which is unnecessary here. Our model represents a dynamical effect where a minimum of two degrees of freedom are necessary, to create two distinct extinction fixed points.

An alternative approach to switching can be found in Filippov systems \cite{filippov1960differential, biak2013some}. In these systems, the switching is neither periodic nor random, but depends on the current state. As such, these systems can be thought of as similar to control-theory, where if a species goes rare, an intervention can help protect it. These systems can, similar to ours, observe switching induced bifurcations in predator-prey systems \cite{dercole2007bifurcation}.

\subsection{Conclusion}

In summary, a species that is eliminated in every static environment can survive once the environment switches between them, provided the switching outpaces the population's collapse. This dynamical route to coexistence is not specific to any model, and we expect it to be relevant well beyond the ecological setting studied here.

\section{Methodology} \label{section:materials-and-methods}

\subsection{Deterministic switching}

All simulations of the dynamics (such as those used in \autoref{fig:vector-fields}D, \autoref{fig:ystar-lambda-vs-alpha}) use a fourth-order Runge Kutta iterator. The analytical results follow the mathematics given below.

The condition for extinction in each environment being can be calculated by straightforwardly finding the fixed points for the system governed by equations \ref{eq:main-equation-dxdt} and \ref{eq:main-equation-dydt} (i.e, setting $\mathrm{d}x/\mathrm{d}t\!=\!0$, and $\mathrm{d}y/\mathrm{d}t\!=\!0$).

Aside from a trivial prey-only fixed point ($y_0^\star\!=\!0$), the system has another fixed point with prey population $x^\star\!=\!\gamma$. As we look for parameter values in which the predator cannot survive, we use the condition that $y^\star\!\leq\! 0$, which gives us \autoref{eq:fixed-environment-death-inequality}.

The fast-switching case simply uses time averaged vector fields, namely

\begin{align}
    \frac{\mathrm{d}x}{\mathrm{d}t} &= \frac{1}{2} \left[ r_1 x \left(1 - \frac{x}{K_1}\right) - x y \right] + \frac{1}{2} \left[ r_2 x \left(1 - \frac{x}{K_2}\right) - x y \right] \label{eq:fast-switching-dxdt} \\
    \frac{\mathrm{d}y}{\mathrm{d}t} &= \frac{1}{2} (x y - \gamma_1 y) + \frac{1}{2} (x y - \gamma_2 y) \label{eq:fast-switching-dydt}
    \end{align}

We can once again calculate the fixed points. This time, we would like for fast switching to keep the predator alive. This requires $y^\star \!>\! 0$, which simplifies to \autoref{eq:switching-environment-persistence-inequality}.

The critical switching rate, $\alpha_c^\text{det}$, can be estimated through Floquet theory. First, we note that \autoref{eq:main-equation-dxdt} for $y\!=\!0$ can be integrated as it is separable:

\begin{equation}
    \int \frac{\mathrm{d}x}{x(1-x/K_E)} = \int r_E\,\mathrm{d}t
\end{equation}

Using partial fractions, 

\begin{equation}
    \ln\!\left(\frac{x}{K_E-x}\right) = r_E t + C.
\end{equation}

The logistic flow on the boundary $\mathcal{X}_E(t, x_\text{in})$, can thus be derived, and is presented in \autoref{eq:logistic-flow-boundary-closed-form}.

After the transient, a system which kills the predator has the prey lie on a periodic boundary orbit $x^\star(t)$ of period $2/\alpha$, where $x^\star(t+2/\alpha)\!=\!x^\star(t)$. If the two `turning points' where the environment changes are $x_a$ and $x_b$, then

\begin{equation}
    x_a = \mathcal{X}_1\!\left(\frac{1}{\alpha};x_b\right) , \quad
    x_b = \mathcal{X}_2\!\left(\frac{1}{\alpha};x_a\right) \label{eq:periodicity-conditions-xaxb}
\end{equation}

As $\mathrm{d}y/\mathrm{d}t \!=\! y(x\! -\! \gamma_E)$, the per-capita growth rate of $y$ is $(x \!-\! \gamma_E)$. The Floquet exponent over one full cycle can be calculated as
\begin{align}
    \lambda(\alpha) &= \frac{\alpha}{2}\int_0^{2/\alpha} \bigl(x_*(t)-\gamma(t)\bigr)\,dt \\
    \begin{split}
    &=\frac{\alpha}{2}\left[\int_0^{1/\alpha}\bigl(\mathcal{X}_1(t;x_b) - \gamma_1\bigr)\,dt \right. \\
    &\left. \qquad +\int_0^{1/\alpha}\bigl(\mathcal{X}_2(t;x_a) - \gamma_2\bigr)\,dt\right] \label{eq:floquet-exponent-derivation-5050}
    \end{split}
\end{align}

Using \autoref{eq:logistic-flow-boundary-closed-form} and combining the logarithmic terms allows us to derive a closed form equation for the Floquet exponent:

\begin{equation} \label{eq:floquet-exponent-closed-form}
    \lambda(\alpha) =\frac{K_1\!+\!K_2\!-\!\gamma_1\!-\!\gamma_2}{2} -\frac{\alpha}{2}\left(\frac{K_1} {r_1}\!-\!\frac{K_2}{r_2}\right) \ln\!\left(\frac{x_a}{x_b}\right)
\end{equation}

The critical switching rate $\alpha_c^\text{det}$ is the solution to the transcendental equation above where $\lambda(\alpha_c^\text{det})\!=\!0$, presented in \autoref{eq:Tc-transcendental-eqn}.

\subsection{Poisson switching}

The analysis follows the same route as the deterministic case, except that the time spent in each environment is no longer the fixed dwell $1/\alpha$, but an exponentially distributed random variable with the same mean $1/\alpha$. The switching rate is the same in both directions, so each environment is occupied half the time on average.

We first record an integral identity for the prey trajectory over a dwell of length $\tau$. Writing the logistic equation \autoref{eq:main-equation-dxdt} on the $y\!=\!0$ boundary as $\mathcal{X}_E \!=\! K_E - (K_E/r_E)\,\mathrm{d}\ln\mathcal{X}_E/\mathrm{d}t$ and integrating from $0$ to $\tau$,

\begin{equation} \label{eq:poisson-integral-identity}
    \int_0^{\tau}\!\mathcal{X}_E(t;x_\text{in})\,\mathrm{d}t = K_E\tau - \frac{K_E}{r_E}\ln\!\left(\frac{\mathcal{X}_E(\tau;x_\text{in})}{x_\text{in}}\right),
\end{equation}
which holds for random $\tau$ as well. Following the deterministic convention, we label the prey density at the start of an environment-1 dwell $x_b$, so that after a dwell $\tau_1$ and a subsequent environment-2 dwell $\tau_2$ the prey is at $x_a \!=\! \mathcal{X}_1(\tau_1;x_b)$ and $x_b' \!=\! \mathcal{X}_2(\tau_2;x_a)$. As $\mathrm{d}\ln\! y/\mathrm{d}t \!=\! x \!-\! \gamma_E$, the predator's log-change over one full cycle is

\begin{equation}
    \Delta\ln y = \int_0^{\tau_1}\!\!\big(\mathcal{X}_1(t;x_b) - \gamma_1\big)\mathrm{d}t + \int_0^{\tau_2}\!\!\big(\mathcal{X}_2(t;x_a) - \gamma_2\big)\mathrm{d}t
\end{equation}

Applying \autoref{eq:poisson-integral-identity} to each dwell and taking expectations (with $\tau_1,\tau_2\overset{\text{iid}}{\sim}\text{Exp}(\alpha)$, so $\mathbb{E}[\tau_i]\!=\!1/\alpha$), then dividing by the expected cycle length $\mathbb{E}[\tau_1+\tau_2]\!=\!2/\alpha$, gives the invasion exponent

\begin{equation} \label{eq:poisson-invasion-exponent-raw}
\begin{split}
    \lambda(\alpha) = \frac{K_1\!+\!K_2\!-\!\gamma_1\!-\!\gamma_2}{2} - \frac{\alpha}{2}\Big(&\frac{K_1}{r_1}\mathbb{E}[\ln(x_a/x_b)] \\[-0.5ex] + &\frac{K_2}{r_2}\mathbb{E}[\ln(x_b^\prime/x_a)]\Big)
\end{split}
\end{equation}

We then consider the stationary state, where $x_b\!=\! x_b^\prime$, the distribution which the prey density at the start of each environment-1 dwell converges to over many cycles. Thus, the prey log-density cannot drift over a cycle, and $\mathbb{E}[\ln(x_b^\prime/x_b)]\!=\!0$. Writing $\ln(x_b^\prime/x_b) \!=\! \ln(x_b^\prime/x_a) + \ln(x_a/x_b)$, this gives $\mathbb{E}[\ln(x_b^\prime/x_a)] \!=\! -\mathbb{E}[\ln(x_a/x_b)]$. Defining the stationary mean log-increment over an environment-1 dwell, $\mathcal{A}(\alpha)\!=\!\mathbb{E}[\ln(x_a/x_b)]$, \autoref{eq:poisson-invasion-exponent-raw} reduces to

\begin{equation} \label{eq:poisson-invasion-exponent-methods}
    \lambda(\alpha) = \frac{K_1\!+\!K_2\!-\!\gamma_1\!-\!\gamma_2}{2} - \frac{\alpha}{2}\left(\frac{K_1}{r_1}\!-\!\frac{K_2}{r_2}\right)\mathcal{A}(\alpha),
\end{equation}
the direct analogue of the deterministic Floquet exponent \autoref{eq:floquet-exponent-closed-form}.

The critical switching rate is the value at which the predator neither grows nor decays, $\lambda(\alpha_c)=0$. Setting \autoref{eq:poisson-invasion-exponent-methods} to zero yields the transcendental equation \autoref{eq:poisson-alphac-transcendental} presented in the main text. There is no closed form, since $\mathcal{A}(\alpha)$ requires the stationary density of the random map $x_b\mapsto x_b' = \mathcal{X}_2(\tau_2;\mathcal{X}_1(\tau_1;x_b))$. We simulate the boundary map by drawing exponential dwell times, iterate it until it reaches its stationary distribution, and then estimate $\mathcal{A}(\alpha)\! =\! \mathbb{E}[\ln(x_a/x_b)]$ as an ensemble average. A comparison to a direct RK4 simulation is shown in \autoref{fig:ystar-lambda-vs-alpha}.

\subsection{General Framework}

We start from assumptions \ref{general-assumption1}, \ref{general-assumption2} and \ref{general-assumption3} listed in the main text. Rather than \ref{general-assumption3}, we consider the following assumption, which follows from assumptions \ref{general-assumption2} and \ref{general-assumption3}, but is a weaker assumption:

\begin{enumerate} \setcounter{enumi}{3}
    \item The species $y$ goes extinct near the boundary in each environment. As the growth of the $y$ when $y\approx 0$ is given by $\partial_y G_i(x,0)$ (linearized from a Taylor expansion), we have $\partial_y G_i(K_i, 0) < 0$.
\end{enumerate}

Without loss of generality, we assume the $K_1 \!<\! K_2$ (the first environment has a fixed point with a lower population). We adopt the notation $f(x) \!=\! F(x,0)$ and $h(x) \!=\! \partial_y G(x, 0)$, and consider the dynamics on the $y\!=\!0$ boundary.

Suppose we construct a periodic switching cycle between bounds $x_a$ and $x_b$, such that $K_1 \!\le\! x_a \!<\! x_b \!\le\! K_2$. As $\mathrm{d}y/\mathrm{d}t \!=\! y\cdot h(x)$, $\mathrm{d} \ln\! y \!=\!  h(x) \mathrm{d}t$. We compute the net logarithmic gain $\Delta \ln y$ over one full cycle. As the net logarithmic gain is $\int_{x_b}^{x_a} h_1(x) dt$ and $\mathrm{d}t \!=\! \mathrm{d}x / f_1 (x)$, we have (in environment 1)

\begin{equation}
    \left[ \Delta \ln y \right]_\textrm{env1} = \int_{x_b}^{x_a} \frac{h_1(x)}{f_1(x)} \mathrm{d}x 
\end{equation}
As $x_a \!<\! x_b$, we reverse the limits to get
\begin{equation}
    \left[ \Delta \ln y \right]_\textrm{env1} = - \int_{x_a}^{x_b} \frac{h_1(x)}{f_1(x)} \mathrm{d}x 
\end{equation}

Similarly, in environment 2, we can derive
\begin{equation}
    \left[ \Delta \ln y \right]_\textrm{env2} = \int_{x_a}^{x_b} \frac{h_2(x)}{f_2(x)} \mathrm{d}x 
\end{equation}

The total gain per cycle can then be written as

\begin{equation}
    \Delta \ln y = \int_{x_a}^{x_b} \underbrace{\left[ \frac{\partial_y G_2(x,0)}{F_2(x,0)} - \frac{\partial_y G_1(x,0)}{F_1(x,0)} \right]}_{R(x)} \mathrm{d}x
\end{equation}

Where, $R(x)$ is the `switching rescue function' described in \autoref{eq:rescue-function} of the main text.

For multiple environments, we note that whatever switching we undergo, the system will still remain in a limit cycle between the smallest and largest carrying capacities. Then, we can consider a switching trajectory between $K_\text{min}\!<\!x_a\!<\!x_b\!<\!K_\text{max}$. This, once again, sweeps over an interval $\Delta x$ twice, once from $x_a\!\to\!x_b$, and once from $x_b\!\to\!x_a$. As $\Delta x\! \to\! 0$, each route is in a constant environment. Thus, if no two constant environments can lead to a positive $\Delta \ln y$ in the interval $\Delta x$, switching cannot rescue the system, though rescue is possible if even one combination of environments can have a positive $R(x)$ for $x\!\in\![K_\text{min}, K_\text{max}]$.

\subsection{The standard viral dynamics model}

We also provide a pharmacological model, to demonstrate the applicability to drug cycling. Consider Nowak and May or Perelson's model for viral dynamics \cite{nowak2000virus, perelson2002modelling}, with uninfected target cells $T$, infected cells $I$, and free viruses $V$:

\begin{align}
    \frac{\mathrm{d}T}{\mathrm{d}t} &= \lambda - d\,T - \beta\,T V \\
    \frac{\mathrm{d}I}{\mathrm{d}t} &= \beta\,T V - \delta\,I \\
    \frac{\mathrm{d}V}{\mathrm{d}t} &= p\,I - c\,V
\end{align}

To draw an analogy to our two-variable case, we assume that the viral dynamics are on a faster timescale (large $c$), and thus, we can assume $\mathrm{d}V/\mathrm{d}t\!=\!0$. We can then derive the quasi-stationary approximation:

\begin{align}
    \frac{\mathrm{d}T}{\mathrm{d}t} &= \lambda - d\,T - \frac{\beta p}{c}\,T I \\
    \frac{\mathrm{d}I}{\mathrm{d}t} &= \frac{\beta p}{c}\,T I - \delta\,I 
\end{align}

The `environments' can be thought of as different drugs, which modulate the parameters. If we define $a\!=\!\beta p/c$, the conditions for switching induced survival are:

\begin{align} \label{eq:nowak-may-perelson-conditions}
\frac{a_1\lambda_1}{d_1\delta_1}<1,\quad
\frac{a_2\lambda_2}{d_2\delta_2}<1,\quad
\frac{(a_1+a_2)(\lambda_1+\lambda_2)}{(d_1+d_2)(\delta_1+\delta_2)}>1
\end{align}

\subsection{Data availability}

The data that support the findings of this article are openly available \cite{rizfn2026githubrepo}.

\section{Acknowledgements}

We thank Kim Sneppen and Bjarke Frost Nielsen for providing valuable feedback on the manuscript. RFN and KK are supported by the Novo Nordisk Fonden Grant No. NNF21OC0065542.

\printbibliography

\appendix

\section{Optimal time in each environment}\label{appendix:alternate-popt}

We study the more general case where, rather than spending its time equally in each environment, the system spends a fraction $p$ of its time in Environment $1$ and $(1-p)$ of its time in Environment $2$ (which we term the `duty cycle'). We characterize the switching by its rate $\alpha$, as in the main text: the full period is $2/\alpha$, of which a fraction $p$ is spent in Environment $1$ (a dwell of $2p/\alpha$) and a fraction $(1-p)$ in Environment $2$ (a dwell of $2(1-p)/\alpha$). At $p\!=\!1/2$ this recovers the equal dwell $1/\alpha$ in each environment used in the main text.

Defining $c_i\! =\! r_i/K_i$, we construct time-averaged parameters:

\begin{align}
    \bar{r}(p) &= p r_1 + (1-p) r_2 \label{eq:dutycycle-timeaveraged-r} \\
    \bar{\gamma}(p) &= p \gamma_1 + (1-p) \gamma_2 \label{eq:dutycycle-timeaveraged-gamma} \\
    \bar{c}(p) &= p \frac{r_1}{K_1} + (1-p) \frac{r_2}{K_2} \label{eq:dutycycle-timeaveraged-c}
\end{align}

We can then derive the condition for survival in the fast-switching case:

\begin{align} \label{eq:duty-cycle-fast-condition}
    \bar{c}(p)\,\bar{\gamma}(p) < \bar{r}(p)
\end{align}

We next turn our attention to the `optimal' duty cycle, i.e, the value of $p$ that maximizes the fast-switching predator population $y^\star$, derived to be

\begin{equation}
    p^{y^\star}_\text{opt} = \frac12 + \frac{\left(r_1-\gamma_1 c_1\right) - \left(r_2-\gamma_2 c_2\right)}{2\left(\gamma_1 - \gamma_2\right)\left(c_1 - c_2\right)} \label{eq:dutycycle-popt-ystar}
\end{equation}

For our previously described case ($r_1=1, r_2=4, K_1=1, K_2=2, \gamma_1=1.1, \gamma_2 = 2.1$), we have $p_\text{opt}^{y^\star} \approx 0.550$.

In addition, we could define the optimal duty cycle in different ways. For example, optimizing for the rate at which the predator grows near the $y=0$ boundary (the invasion exponent), gives us

\begin{equation}
    p_\text{opt}^\lambda = \frac{1}{c_1 - c_2}\left(\sqrt{\frac{r_1 c_2 - r_2 c_1}{\gamma_1 - \gamma_2}} - c_2\right)
\end{equation}

For the same parameters, this gives a distinct value of $p_\text{opt}^\lambda \approx 0.586$.

\begin{figure}
    \centering
    \includegraphics[width=\linewidth]{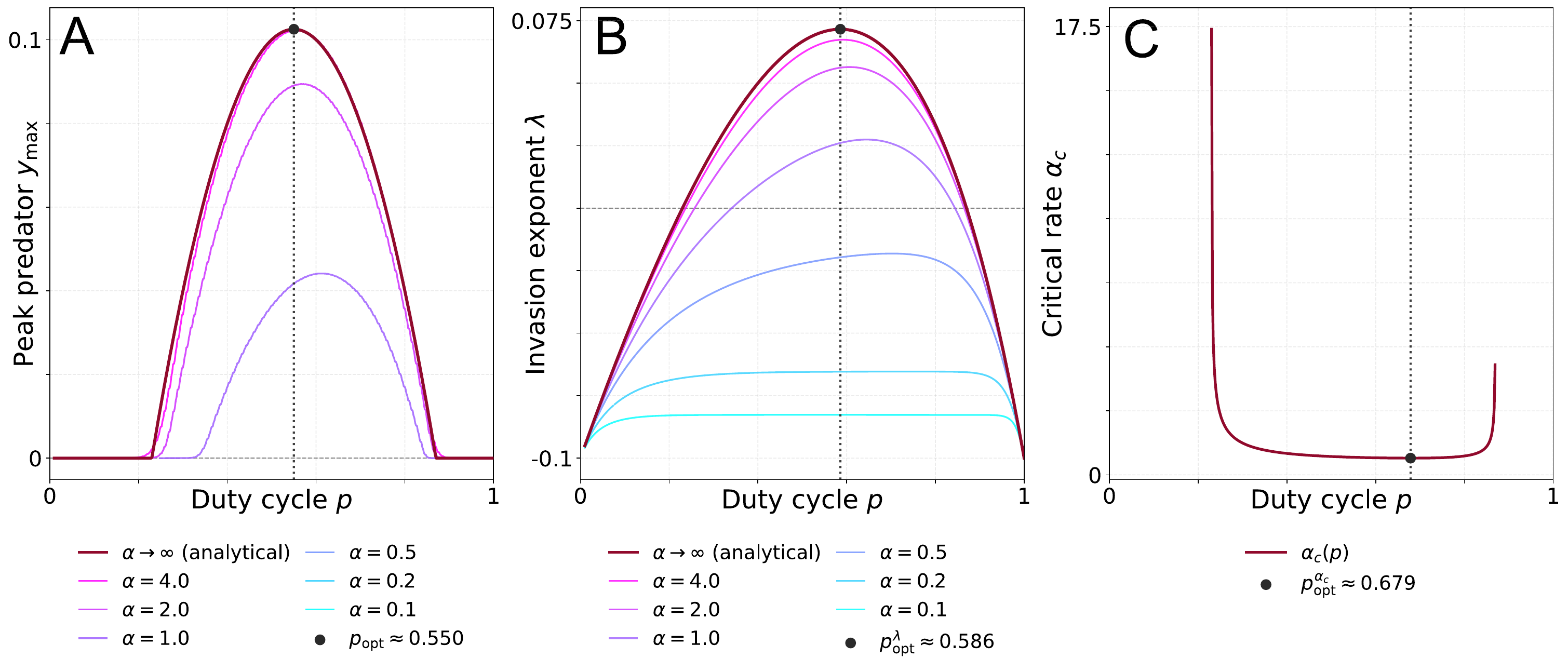}
    \caption{Optimal duty cycles, for varying targets. A: Maximizing the steady-state predator peak, $\max(y^\star)$, for different switching rates (estimated numerically from simulations at finite $\alpha$). B: Maximizing the invasion exponent $\lambda$, the strength at which $y$ grows from $y\!\approx\! 0$. C: The critical switching rate $\alpha_c$ as a function of $p$. All results use parameters $r_1\!=\!K_1\!=\!1$, $\gamma_1\!=\!1.1$, $r_2\!=\!4$, $K_2\!=\!2$, $\gamma_2\!=\!2.1$.}
    \label{fig:duty-cycles-popt}
\end{figure}

The critical switching rate $\alpha_c$ also depends on the duty cycle. Optimizing for a cycle with the minimum $\alpha_c$ (as the predator survives wherever $\alpha\!>\!\alpha_c$, this finds the duty cycle most robust to changes in the switching frequency) gives us the transcendental equation

\begin{equation}
    \begin{split} \label{eq:duty-cycle-Tc-transcendental-eqn}
        (K_1 - \gamma_1)p_\text{opt}^{\alpha_c} + (K_2-\gamma_2)(1-p_\text{opt}^{\alpha_c}) \\ =  \frac{\alpha_c}{2} \left[\frac{K_1}{r_1}\ln\!\left(\frac{x_a}{x_b}\right) + \frac{K_2}{r_2}\ln\!\left(\frac{x_b}{x_a}\right) \right],
    \end{split}
\end{equation}

where $x_a$ and $x_b$ depend on $\alpha_c(p)$ and $p$ via the boundary flow \autoref{eq:si-logistic-flow} and

\begin{equation} \label{eq:tc-popt-cyclic-xa*-xb*-lc}
    x_a = \mathcal{X}_1\!\left(\frac{2p_\text{opt}^{\alpha_c}}{\alpha_c};\, x_b\right), \quad x_b = \mathcal{X}_2\!\left(\frac{2(1-p_\text{opt}^{\alpha_c})}{\alpha_c};\, x_a\right)
\end{equation}

The three quantities and their optima are displayed in \autoref{fig:duty-cycles-popt}.

\section{Derivations for \autoref{appendix:alternate-popt}}

We assume the system spends a fraction of time $p$ in environment 1, and a fraction $(1-p)$ in environment 2. As in the fast-switching case of the main text, the dynamics are governed by the time-averaged vector fields, now weighted by $p$ and $(1-p)$ rather than equally. We define $c = r/K$, and using the time-averaged parameters defined in Equations \ref{eq:dutycycle-timeaveraged-r}, \ref{eq:dutycycle-timeaveraged-gamma} and \ref{eq:dutycycle-timeaveraged-c}, our system reduces to

\begin{align}
\frac{dx}{dt} &= \bar{r}(p) x - \bar{c}(p) x^2 - x y \\
\frac{dy}{dt} &= y \big( x - \bar{\gamma}(p) \big)
\end{align}

Solving for the fixed points and using the condition that $y^\star>0$ gives us \autoref{eq:duty-cycle-fast-condition}.

The steady-state predator density is given by

\begin{equation}
    y^\star(p) = \bar{r}(p) - \bar{\gamma}(p) \bar{c}(p)
\end{equation}

Differentiating with respect to $p$ would give us the optimal duty cycle that maximizes $y^\star$.

We introduce new variables, which capture the change between the environments ($\Delta r = r_1 - r_2$): 
\begin{align}
    \bar{r}(p) &= p \Delta r + r_2 \implies \frac{\mathrm{d}\bar{r}(p)}{\mathrm{d}p} = \Delta r \\
    \bar{\gamma}(p) &= p \Delta \gamma + \gamma_2 \implies \frac{\mathrm{d}\bar{\gamma}(p)}{\mathrm{d}p} = \Delta \gamma \\
    \bar{c}(p) &= p \Delta c + c_2 \implies \frac{\mathrm{d}\bar{c}(p)}{\mathrm{d}p} = \Delta c
\end{align}

Using these, we can derive

\begin{equation}
    p_\text{opt}^{y^\star} = \frac{\Delta r - c_2 \Delta \gamma - \gamma_2 \Delta c}{2 \Delta \gamma \Delta c},
\end{equation}
which can be reduced to \autoref{eq:dutycycle-popt-ystar}.

In the fast switching case, as $1/y \cdot\mathrm{d}y/\mathrm{d}t = \mathrm{d}(\ln\!y)/\mathrm{d}t = \lambda_\text{fast}$, we can write

\begin{equation} \label{eq:lambda-fast-dutycycle}
    \lambda_\text{fast}(p) = \frac{\bar{r}(p)}{\bar{c}(p)} - \bar{\gamma}(p)
\end{equation}
 
To find $p_\text{opt}^\lambda$, we set $\mathrm{d}\lambda_\text{fast}/\mathrm{d}p = 0$, and get
 
\begin{equation}
    p_\text{opt}^\lambda=\frac{1}{c_1-c_2}\left(\sqrt{\frac{r_1 c_2-r_2 c_1}{\gamma_1-\gamma_2}}-c_2\right)
\end{equation}

As $\alpha_c$ will depend on $p$, we can calculate $p_\text{opt}^{\alpha_c}$, where $\alpha_c$ is minimized.

The dynamics of $x$ in a fixed environment are given, as in the main text, by the logistic flow on the $y=0$ boundary,

\begin{equation} \label{eq:si-logistic-flow}
     \mathcal{X}_E(t; x_{\mathrm{in}}) = \frac{K_E x_{\mathrm{in}} e^{r_E t}}{K_E + x_{\mathrm{in}}(e^{r_E t} - 1)}
\end{equation}

Again, if we have a limit cycle where the prey moves from $x_b$ to $x_a$ in the first wave, and from $x_a$ to $x_b$ in the second, we now have

\begin{equation} \label{eq:popt-cyclic-xa*-xb*-lc}
    x_a = \mathcal{X}_1\big(\tfrac{2p}{\alpha};\, x_b\big), \qquad x_b = \mathcal{X}_2\big(\tfrac{2(1-p)}{\alpha};\, x_a\big)
\end{equation}

We can then derive a similar equation to the main-text invasion-exponent analysis, as

\begin{align}
\begin{split}
    \lambda (\alpha, p) &= \frac{\alpha}{2} \left[ \int_0^{2p/\alpha} \left(\mathcal{X}_1(t;\,x_b) - \gamma_1 \right) dt \right. \\
    &\quad \left. + \int_0^{2(1-p)/\alpha} \left(\mathcal{X}_2(t;\,x_a) - \gamma_2 \right) dt \right]
\end{split} \\[2ex]
\begin{split}
    &= (K_1 - \gamma_1)p + (K_2-\gamma_2)(1-p) \\
    &\qquad - \frac{\alpha}{2} \left[\frac{K_1}{r_1}\ln\!\left(\frac{x_a}{x_b}\right) + \frac{K_2}{r_2}\ln\!\left(\frac{x_b}{x_a}\right) \right]
\end{split}
\end{align}

As the critical switching rate $\alpha_c$ is defined where $\lambda(\alpha_c, p)=0$, \autoref{eq:duty-cycle-Tc-transcendental-eqn} is derived.

\section{The `shielding' model}\label{appendix:shielding-model}

The Lotka-Volterra setup requires three parameters ($r$, $K$, $\gamma$) to all be independently changed in order to observe the switching-induced persistence. Below, we present the `shielding model', which requires only two parameters, $K$ and $\mu$:

\begin{align}
    \frac{\mathrm{d}x}{\mathrm{d}t} &= x\left(1-\frac{x}{K}\right) - x y e^{-\mu x}  \label{eq:shielding-model-dxdt} \\
    \frac{\mathrm{d}y}{\mathrm{d}t} &= x y e^{-\mu x} - y \label{eq:shielding-model-dydt}
\end{align}

The fundamental change is in the predation dynamics. When $\mu>0$, the predator's efficacy at targeting the prey goes down as the prey population grows. On the macroscale, this can be thought of as bison forming a herd, making them harder to be attacked by predators. On the microscale, it could correspond to spatial dynamics, such as where a phage can only attack a colony of bacteria on the surface, despite the colony growing in the bulk \cite{eriksen2018growing}.

The same fixed-point analysis applies, and we can calculate the conditions for switching-induced persistence to be (where $\tilde{K} = 2K_1 K_2/(K_1+K_2)$)

\begin{equation} \label{eq:shielding-conditions}
        K_1 e^{-\mu_1 K_1} \leq 1, \quad K_2 e^{-\mu_2 K_2} \leq 1, \quad \tilde{K}\,\frac{e^{-\mu_1 \tilde{K}}\!+\!e^{-\mu_2 \tilde{K}}}{2} > 1
\end{equation}

For instance, these conditions hold for the parameters $K_1\!=\!1, \mu_1\!=\!0, K_2\!=\!2, \mu_2\!=\!0.4$.

\newpage

\section{Critical $\alpha$ scaling}

\begin{figure}[h!]
    \centering
    \includegraphics[width=\linewidth]{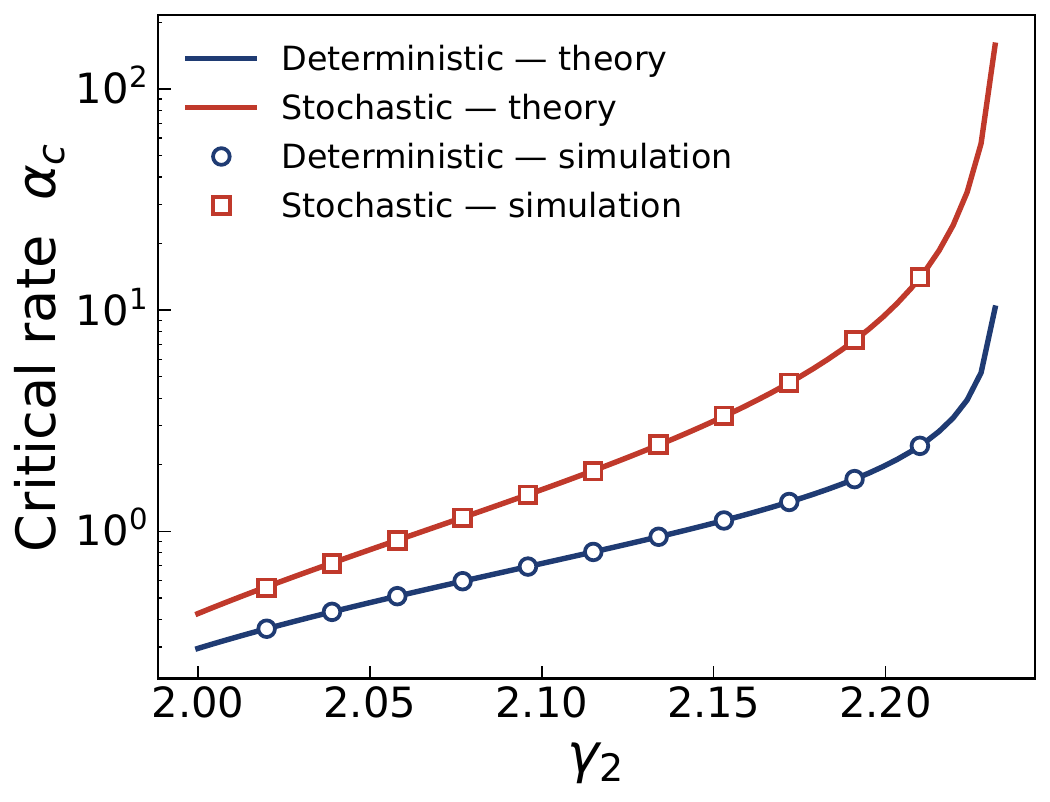}
    \caption{The critical switching rate $\alpha$ versus $\gamma_2$ under both deterministic (red) and stochastic (blue) switching, for $K_1\!=\!1, r_1\!=\!1, \gamma_1\!=\!1.1, K_2\!=\!2, r_2\!=\!4$. The markers represent data from simulations, while the curves are theoretically derived where the turning points $x_a$ and $x_b$ are obtained by fixed point iteration of the flow. The stochastic system needs to switch faster to survive, as bad luck could lead to a long period in one environment where the system falls into the $y=0$ attractor.}
    \label{fig:stochastic-critical-alpha}
\end{figure}

\onecolumn

\section{Generality of $R(x)$}

\begin{figure*}[h!]
    \centering
    \includegraphics[width=\linewidth]{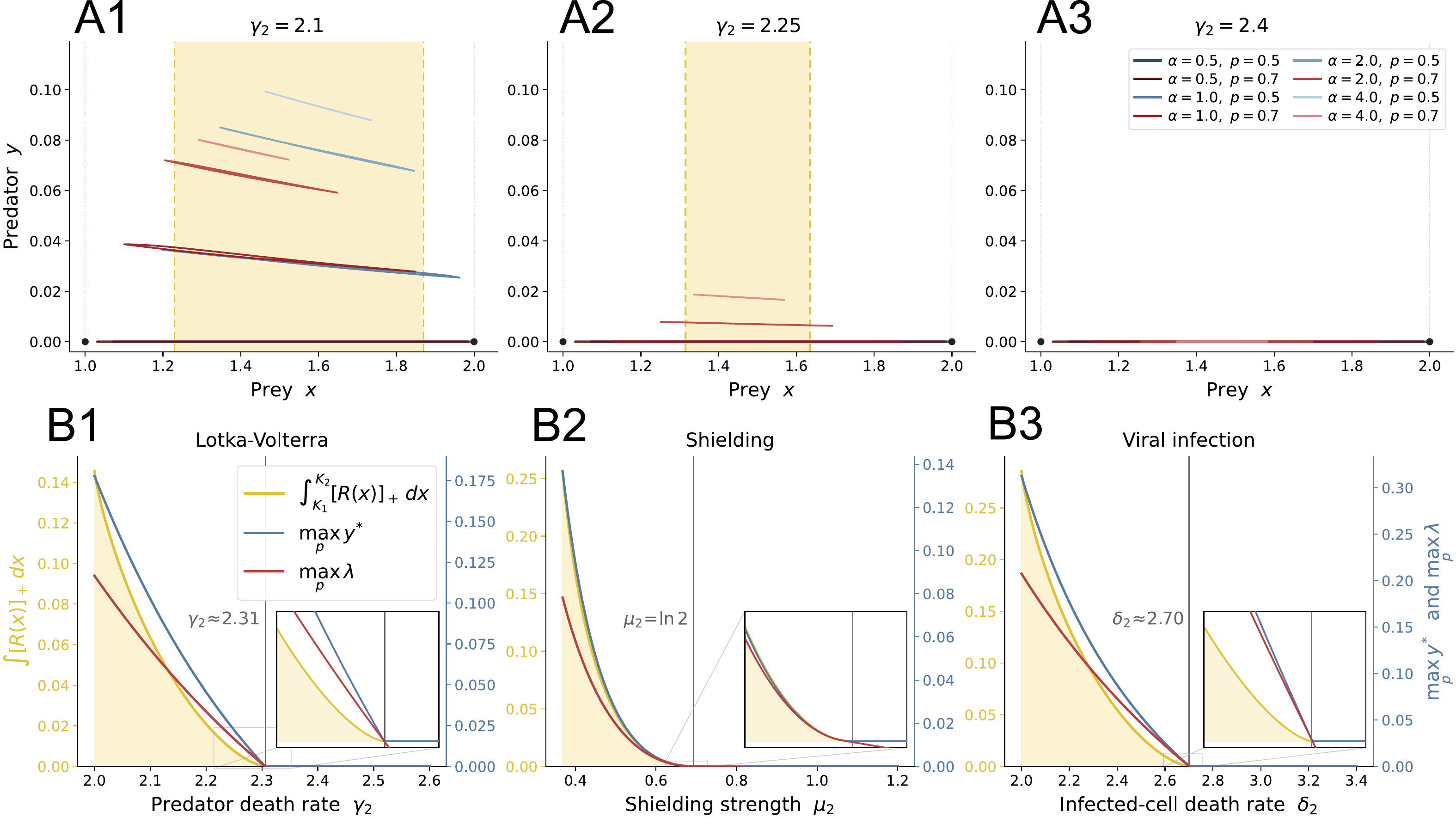}
    \caption{A: The limit cycles for obtained for varying values of $\gamma_2$, with  $r_1\!=\!K_1\!=\!1$, $\gamma_1\!=\!1.1$, $r_2\!=\!4$, $K_2\!=\!2$. A1: $\gamma_2\!=\!2.1$, A2: $\gamma_2\!=\!2.25$, A3: $\gamma_2\!=\!2.4$. Blue lines represent a duty cycle of $p\!=\!0.5$, red represent $p\!=\!0.7$. The shaded yellow region marks the area in phase space where $R(x)$ has positive values. When $R(x)$ is positive across a large region, limit cycles can exist for both duty cycles. As $\gamma_2$ increases, $p\!=\!0.5$ systems go extinct, but as $R(x)$ is positive, there has to be some limit cycle that can exist, which can be seen in the $p\!=\!0.7$ case. Once $R(x)$ is negative across the interval, switching cannot rescue the system. B1: Plots of the integral over the positive parts of $R(x)$ (yellow), the max $y^\star$ in fast switching (blue) and the max invasion exponent $\lambda$ (red). Note that the values of $p$ that maximize the two do not need to be the same. Nevertheless, they can be seen to go to 0 together, at $\gamma_2\!\approx\!2.3$. B2: The same plots for the shielding model (\autoref{appendix:shielding-model}), where $K_1\!=\!1$, $\mu_1\!=\!0$, $K_2\!=\!2$, and $\mu_2$ is varied. B3: The same for the Nowak-May/Perelson viral dynamics model \cite{nowak2000virus, perelson2002modelling}, for $d_1\!=\!\lambda_1\!=\!a_1\!=\!1, \delta_1\!=\!1.2, d_2\!=\!4, \lambda_2\!=\!8, a_2=1$ and $\delta_2$ varies. Again, $R(x)$ predicts the regime where switching can rescue the system.}
    \label{fig:Rfunction-limitcycles}
\end{figure*}

\end{document}